# Reti bayesiane per lo studio del fenomeno degli incidenti stradali tra i giovani in Toscana

*Filippo Elba[1], Lisa Gnaulati[2], Fabio Voeller[3]*

**SOMMARIO**

È qui affrontato il problema degli incidenti stradali degli adolescenti toscani alla guida. Lo si fa partendo dalle informazioni della banca dati Edit dell'Osservatorio di Epidemiologia della Toscana.
La complessità ed eterogeneità dei dati a disposizione costituiscono un campo interessante per l'applicazione di metodi di apprendimento automatico. Per questo, viene qui proposta un'analisi attraverso l'utilizzo di una rete probabilistica bayesiana, utile per individuare le relazioni tra le variabili osservate e i comportamenti e le caratteristiche dei giovani che in maniera più significativa si associano con una guida spregiudicata.
In prospettiva, partendo da questa rete, se ne potrebbe realizzare una causale, uno strumento in grado di limitare questo tragico fenomeno.

**Parole chiave**: incidenti stradali, adolescenti, reti bayesiane, apprendimento automatico.

**ABSTRACT**

This paper aims to analyse adolescents' road accidents in Tuscany. The analysis is based on the database Edit of *Osservatorio di Epidemiologia della Toscana*.
Complexity and heterogeneity of Edit's data represent an interesting scope to apply machine learning methods. In particular, in this paper is proposed an analysis based on a Bayesian probabilistic network, used to discover relationships between adolescents' characteristics and behaviours that are more often associated with an audacious driving style.
The probabilistic network developed by this study can be considered a useful starting point for follow-up reasearches, aiming to develop a causal network, a tool to limit this phenomenon.

**Key words**: road accidents, adolescents, Bayesian networks, machine learning.

## 1.     GLI INCIDENTI STRADALI DEGLI ADOLESCENTI

### 1.1 Gli incidenti stradali in Italia: costi e statistiche

Gli incidenti stradali sono un problema di assoluta priorità per la collettività e per la sanità pubblica. Essi rappresentano, a livello mondiale, la prima causa di morte nei giovani di età tra i 15 e i 19 anni e la seconda nei ragazzi di 10-14 anni e 20-24 anni. Ai costi sociali e umani si aggiungono anche quelli economici, calcolati tra l'1 ed il 3% del PIL di ciascun paese, tanto da porre la questione della sicurezza stradale all'attenzione dell'agenda di tutti i sistemi sanitari (Ars Toscana, 2015).

---

[1] Assistente alla ricerca e cultore della materia (Università degli Studi di Firenze), e-mail: filippo.elba@unifi.it.
[2] Ricercatrice (Osservatorio di Epidemiologia - ARS Toscana), e-mail: lisa.gnaulati@ars.toscana.it.
[3] Dirigente di ricerca (Osservatorio di Epidemiologia - ARS Toscana), e-mail: fabio.voller@ars.toscana.it.

Secondo i dati dell'Istat riferiti al 2014, su 3.381 vittime di incidenti stradali in Italia, ca. 200 sono nella fascia adolescenziale (10-19 anni) (di questi, una trentina sono pedoni). Interessante notare come, tra i soggetti di età compresa tra 15 e 19 anni, il tasso di mortalità per incidenti stradali è pari a 60 ogni 100 mila unità, di poco al di sopra della media nazionale, ferma a 55,6. Su 251.147 feriti, sono, invece, quasi 24 mila quelli nella fascia d'età 14-19 anni (circa 2.500 sono pedoni. Nei restanti, come nel caso delle vittime, sono ricompresi anche eventuali passeggeri e ciclisti, quindi non necessariamente soggetti alla guida di un mezzo motorizzato).

Sulla base di stime ministeriali elaborate nel 2015 è possibile quantificare il costo di questi eventi, a prescindere dall'età dei soggetti incidentati. In maniera generale, i costi di un sinistro possono essere classificati in:

- umani, riferiti alle vittime di incidente stradale e derivati dalla perdita di produttività per la società, dalla perdita affettiva, dolore e sofferenza delle persone coinvolte e dei parenti delle vittime, dai costi delle cure mediche cui sono sottoposte le vittime;
- generali, riferiti all'incidente stradale, derivanti dai danni al veicolo, dalle spese per il rilievo degli incidenti da parte delle forze di polizia e dei servizi di emergenza, dai costi legali e amministrativi di gestione, dai danni causati all'infrastruttura stradale e agli edifici.

Partendo da questa tassonomia, il Ministero, per il 2010, calcola i seguenti costi medi per tipologia di incidente, in base alle lesioni riportate dalle vittime (Tab. 1.1):

**Tabella 1.1: costi degli incidenti stradali per livello di gravità - Anno 2010 (valori in mln di €)**

| Gravità | Costo medio umano in base alla gravità delle lesioni riportate | Costo medio incidente in base alla gravità dell'incidente stradale |
| --- | --- | --- |
| **Mortale** | 1,503 | 1,642 |
| **Con lesioni gravi** | 0,197 | 0,309 |
| **Con lesioni lievi** | 0,017 | 0,032 |
| **Con lesioni (senza distinguere in base alla gravità)** | 0,042 | - |

*Fonte:* Ministero delle Infrastrutture e dei Trasporti.

Alla luce di questi dati, con riferimento al 2013, il costo nazionale stimato riferito ai soli sinistri stradali con danni alle persone (morti o lesioni, più o meno gravi) è così quantificato (Tab. 1.2):

**Tabella 1.2: costo sociale totale dell'incidentalità con danni a persone - Anno 2013**

| | |
| --- | --- |
| **Costo totale dei decessi** | 5.115.069.990 |
| • *Costo medio umano per decesso (€)* | 1.503.990 |
| • *N° morti* | *3.401* |
| **Costo totale dei feriti** | 10.896.428.367 |
| • *Costo medio umano per ferito (€)* | 42.219 |
| • *N° feriti* | *258.093* |
| **Costi Generali Totali** | 1.995.716.760 |
| • *Costi Generali medi per incidente (€)* | 10.986 |
| • *N° incidenti stradali* | *181.660* |
| **Costo sociale totale - Italia** | **18.007.215.117** |

*Fonte:* Ministero delle Infrastrutture e dei Trasporti.

I 18,01 mld di € lievitano a 24,34 mld di € quando si considerano anche gli incidenti senza danni a persone.

Riuscire a individuare gli elementi utili per poter fare prevenzione innanzitutto tra i giovani è fondamentale per salvare vite umane e, in secondo luogo, per ridurre i costi a carico della collettività.

**1.2 La letteratura in materia di incidenti stradali degli adolescenti**

Nonostante l'importanza del fenomeno qui analizzato, piuttosto scarsa è la letteratura al riguardo. Tra i pochi lavori, si segnala quello di Elvik e Lund (2015): i due autori sottolineano come, nonostante dagli anni '70 in poi il tasso complessivo di incidenti fatali si sia notevolmente ridotto, tra i giovani maschi la situazione è rimasta sostanzialmente costante. Lo studio individua le ragioni di ciò in fattori biologici (ormoni, sviluppo celebrale), nell'eccessiva autostima ed in un atteggiamento di ribellione e di sfida dei limiti tipico di questa età. Nonostante il limitato margine di manovra rispetto a questi elementi, è evidenziata, tuttavia, una certa efficacia dei meccanismi di ricompensa a favore di chi mantiene una velocità limitata.

Rahim Khan et al. (2015) propongono uno studio globale sull'andamento del tasso di incidenti stradali di bambini e adolescenti[4] (1-19 anni). Essi guardando a quello che succede tra 1990 e 2013, raggruppando i Paesi oggetto di analisi in base al reddito ed all'area geografica. In un certo senso, dunque, viene messo l'accento sui ruoli che il contesto socio economico e geografico giocano nel determinare il fenomeno. Emerge come, a livello complessivo, si assista ad una generale riduzione dei tassi relativi ad incidenti, più marcata nei Paesi ad alto reddito, meno in quelli a medio e basso reddito. Unica eccezione è rappresentata dai Paesi a basso reddito dell'Africa Sub-Sahariana.

La restante letteratura in materia si limita, per lo più, a presentare statistiche descrittive riferite al problema, senza indagare approfonditamente sui motivi che possono esserci alla base. A tal proposito si segnala il report multimediale dell'Organizzazione Mondiale della Salute (2014) "Health for The World's Adolescence: A Seconde Chance in The Second Decade", che esamina, a tutto tondo, diversi aspetti riguardanti la salute degli adolescenti, tra cui gli incidenti stradali.

**1.3 La banca dati Edit 2015[5]**

La banca dati Edit 2015, frutto dell'indagine su "Epidemiologia dei determinanti dell'infortunistica stradale in Toscana", costituisce la quarta rilevazione effettuata dall'Osservatorio di Epidemiologia dell'Ars Toscana nell'ambito di un progetto che si propone come punto di riferimento, in Toscana e in Italia, per la produzione di analisi e riflessioni che consentano di migliorare conoscenza e capacità di intervento sulla realtà dell'infortunistica stradale degli adolescenti e delle sue determinanti.

La rilevazione, ideata e realizzata a partire dal 2005, è ripetuta negli anni 2008, 2011 e 2015. La lista e la composizione delle classi da inserire nel campione è fornita dal Provveditorato degli Studi regionale. Il campione estratto è stratificato per Asl e per tipologia di Istituto scolastico superiore. Per ottenere dei risultati rappresentativi a livello di Asl, sono selezionati 400 soggetti circa per ogni territorio, corrispondenti a circa 4 scuole, ad eccezione della Asl-Firenze dove, per ragioni legate alla dimensione demografica, ne sono selezionate 11. Per ogni Asl, gli istituti sono estratti con campionamento sistematico, con probabilità di estrazione proporzionale al numero di studenti per istituto, previo ordinamento della lista degli istituti per tipologia. Per ogni scuola arruolata nello studio sono sorteggiate cinque classi, dalla I alla V, appartenenti a sezioni diverse.

---

[4] Sono aggregati i dati riferiti sia a vittime "vulnerabili" (pedoni, ciclisti, motociclisti), sia "non vulnerabili" (automobilisti).
[5] Le informazioni qui esposte sono tratte dal documento di sintesi dell'indagine Edit 2015 (https://www.ars.toscana.it/files/edit/2015/sintesi_EDIT_2015_fronte3.pdf).

L'indagine 2015, realizzata nei quattro mesi tra febbraio e maggio, coinvolge 5.077 studenti appartenenti a 57 istituti secondari della Toscana, di età compresa tra i 14 e i 19 anni: il 54,2% maschi e il 45,8% femmine. Le età maggiormente rappresentate sono 16 e 17 anni (rispettivamente 1.007 e 1.003 alunni), seguite da 15 e 18 anni (rispettivamente 936 e 904 alunni), mentre frequenze più basse sono registrate per i quattordicenni (674 casi) e per i diciannovenni (553). La distribuzione delle classi, invece, fa registrare le frequenze maggiori nelle prime (1.132 alunni intervistati), seguite dalle terze (1.043 alunni), dalle seconde, dalle quarte (circa 1.000 alunni entrambe) e infine dalle quinte (899 alunni).

L'ultima indagine tocca temi quali: i comportamenti alla guida; i rapporti con i pari e con la famiglia; l'andamento scolastico; l'attività sportiva; i comportamenti alimentari; i consumi di bevande alcoliche e tabacco; l'uso di sostanze stupefacenti; i comportamenti sessuali e il fenomeno del bullismo; lo stato emotivo; la propensione al gioco d'azzardo; la qualità del sonno dei ragazzi (novità rispetto alle rilevazioni precedenti).

Tra le evidenze che emergono nel 2015 rispetto al passato, vi sono la riduzione del consumo di tabacco, di frutta e verdura e della popolazione in sovrappeso. Resta stabile la quota di chi pratica regolarmente attività fisica, mentre crolla la percentuale di chi gioca d'azzardo. Molti di questi fenomeni sembrano una chiara e diretta conseguenza delle ridotte possibilità economiche delle famiglie e, pertanto, anche dei loro figli. Si tratta di effetti paradossali della crisi economica, che nel breve periodo contribuiscono a migliorare o a limitare le ripercussioni di alcuni comportamenti a rischio. Restano, però, alcune criticità, solo parzialmente riconducibili agli effetti sulla salute, stavolta negativi, della crisi: aumentano gli sperimentatori di sostanze (consumatori almeno una volta nella vita) anche se rimangono sostanzialmente stabili i consumatori abituali; aumenta lo stato di "distress"[6] dei giovani toscani; si conferma uno stile di consumo alcolico molto globalizzato che si concentra durante i weekend con bevande a più alta gradazione alcolica, quasi mai vino, e che si caratterizza per frequenti episodi eccedentari (ubriacature).

Una considerazione a parte merita il continuo aumento dei ragazzi che mettono in pratica comportamenti sessuali non protetti che, nel 2015, sono arrivati a rappresentare poco più della metà di coloro che hanno avuto almeno un rapporto sessuale completo. Questo rimette prepotentemente in campo la necessità di informare i ragazzi rispetto a temi fondamentali, come l'uso del profilattico al fine di prevenire una malattia a trasmissione sessuale piuttosto che come anti concezionale.

Di notevole interesse anche i dati sulla qualità del sonno: quasi un terzo del campione dorme meno di 7 ore a notte, quando le raccomandazioni suggerirebbero che, sotto i 20 anni, sarebbero necessarie almeno otto ore e mezza per notte; quasi un quarto del campione femminile dichiara di avere un sonno disturbato; solo il 40% del campione totale dichiara di avere un sonno profondo.

Secondo un modello di regressione logistica sviluppato all'interno del report 2015, volto ad individuare i fattori che più significativamente si associano agli incidenti stradali degli adolescenti, il rischio più basso di incidente è associato a guidare in condizioni di stanchezza. La guida dopo aver bevuto troppo si rivela un fattore significativo ma al di sotto delle aspettative, in linea con un fattore di rischio che in letteratura riveste un ruolo minore, vale a dire fumare una sigaretta durante la guida. Ascoltare musica ad alto volume e guidare dopo aver assunto sostanze stupefacenti sono mediamente rilevanti. La variabile del modello che più di ogni altra aumenta il rischio di incorrere in un sinistro è la guida in condizioni di ritardo.

Infine, su un campione ristretto di studenti (1.017 unità, pari al 20% ca. del campione iniziale) con dati completi per quel che concerne le informazioni su anagrafica e comportamenti a rischio (per esempio, gioco d'azzardo, consumo di tabacco, uso di cannabis e di altre droghe nell'ultimo mese, etc.) è realizzata un'analisi dei cluster che porta all'individuazione di 3 macro gruppi di soggetti, così definiti:

---

[6] Lo studio Edit, in linea con alcune ricerche internazionali svolte sulla stessa fascia di popolazione, ha scelto di utilizzare la "Kessler Psychological Distress Scale" (K6). Questo strumento, misurando le condizioni mentali su di una lista di sintomi di malessere psicologico auto riferito, consente di identificare condizioni specifiche di disagio psicologico, definito distress.

- "emancipate", costituito da 269 unità (26,5% del totale), prevalentemente femmine minorenni, che rivelano la percentuale più alta di consumo di tabacco, hanno rapporti mediocri o poco buoni con la propria famiglia e hanno ripetuto almeno un anno scolastico. Si tratta di un gruppo che rivela molte virtù: dai bassi rischi durante la guida alla pratica di attività sportiva, fino ad arrivare ad un consumo appropriato di frutta e verdura;
- "eccedentari", costituito da 473 record (46,5% del totale), prevalentemente maschi maggiorenni che, per quasi tutti i comportamenti a rischio considerati, fanno registrare valori più alti rispetto agli altri due gruppi, denotando uno stile di vita molto sregolato. Nello specifico è questo gruppo a rivelare il maggior uso e abuso di alcol, droghe e tabacco. È, inoltre, elevata la percentuale di incidenti con conseguenze tali da richiedere cure mediche e di comportamenti alla guida rischiosi. Sempre questo gruppo riferisce frequenze elevate per quanto riguarda gli atteggiamenti da bullo e l'aver avuto il primo rapporto sessuale in età precoce (11-14 anni);
- "moderati", costituito da 270 unità (26,5% del totale), prevalentemente maschi minorenni che, per quasi tutti i comportamenti a rischio analizzati, fanno registrare i valori più bassi rispetto agli altri due gruppi. Lo stile di vita di questi ragazzi suggerisce un disinteresse ad eccedere, dovuto evidentemente alla giovane età. Le uniche due criticità che presentano sono: il basso consumo di frutta e verdura; l'elevato numero di coloro che rivelano di aver avuto i primi rapporti sessuali in età precoce.

## 2. IL "BAYESIANESIMO"

La statistica bayesiana si contraddistingue per la centralità e la rilevanza dell'incertezza e per un approccio soggettivo rispetto a quello, almeno apparentemente, più oggettivo dei frequentisti. L'utilizzo del linguaggio delle probabilità caratterizza questo paradigma inferenziale.

**2.1 Il teorema di Bayes**

Alla base di questo approccio inferenziale vi è il teorema di Bayes[7]. Esso afferma:

$$P(A|B) = \frac{P(B|A)P(A)}{P(B)} \quad (2.1)$$

La probabilità che si verifichi il fenomeno A, condizionatamente a B, è pari al prodotto tra la probabilità che si verifichi il fenomeno B condizionato ad A e la probabilità che si verifichi il fenomeno A, diviso la probabilità che si verifichi il fenomeno B.
Una lettura più generale del teorema di Bayes porta ad una visione della statistica diversa da quella frequentista. Si riconsideri la (2.1) interpretando *A* come "ipotesi" e *B* come "dati": la probabilità a posteriori $P(A|B)$, vale a dire il giudizio riferito a delle ipotesi condizionato ai dati raccolti, è uguale alla probabilità a priori $P(A)$, ossia il contenuto delle ipotesi iniziali, per la verosimiglianza $P(B|A)$, cioè il contenuto emerso dalla valutazioni dei dati condizionati alle ipotesi inziali. Il tutto diviso per $P(B)$, il normalizzatore (Domingos, 2015). L'approccio bayesiano, "soggettivista", permette, dunque, allo studioso di fare delle assunzioni a priori, basate sul suo stato informativo, che entrano direttamente nel modello e che possono, poi, essere rafforzate, rigettate o corrette sulla base dell'informazione contenuta nei dati.

---

[7] Il teorema compare in un'opera postuma di Thomas Bayes, morto nel 1761, dal titolo "*An Essay towards solving a Problem in the Doctrine of Chances*". A pochi anni di distanza, del tutto all'oscuro del lavoro dell'autore inglese, il matematico Laplace arrivò alle medesime conclusioni.

Strettamente connesso alla (2.1) è il concetto di probabilità condizionata:

$$P(A|B) = \frac{P(A \cap B)}{P(B)} \quad (2.2)$$

La probabilità che si verifichi un evento *A* condizionatamente ad un altro evento *B* è pari all'intersezione tra i due eventi *A* e *B*, fratto la probabilità relativa all'evento *B* condizionante. Stessa cosa:

$$P(B|A) = \frac{P(A \cap B)}{P(A)} \quad (2.3)$$

Linearizzando la (2.3), si ottiene:

$$P(A \cap B) = P(B|A)P(A) \quad (2.4)$$

Sostituendo la (2.4) alla (2.2), si ritorna alla (2.1).
Quanto appena visto è alla base della cosiddetta "chain rule": si può calcolare le probabilità riferita ai singoli eventi che costituiscono una distribuzione congiunta usando le probabilità condizionali.
Considerando *n* eventi, la (2.4) può essere così generalizzata:

$$P(A_1 \cap A_2 \dots A_n) = P(A_n|A_{n-1}, \dots, A_1) * P(A_1 \cap A_2 \dots A_{n-1}) \quad (2.5)$$

Ripetendo la scomposizione per ogni evento, si giunge alla:

$$P\left(\bigcap_{i=1}^{n} A_n\right) = \prod_{i=1}^{n} P\left(A_i | \bigcap_{j=1}^{i-1} A_j\right) \quad (2.6)$$

Sfruttando le proprietà della chain rule nascono, alla fine degli anni '80, le reti bayesiane. Il primo a parlarne è Pearl (1985). Pearl (1988) e Neapolitan (1989) definiscono, negli anni successivi, i diversi aspetti riguardanti questo strumento, facendone un nuovo oggetto di studi.

**2.2 Le reti bayesiane**

La rete bayesiana è una rappresentazione grafica delle relazioni di dipendenza tra le variabili di un sistema. Essa è utilizzata per individuare le relazioni di indipendenza, assoluta e condizionale, tra variabili, al fine di ridurre il numero delle combinazioni delle variabili da analizzare, sulla base della chain rule. L'utilizzo di questo tipo di strumento è tanto più vantaggioso quanto più alto è il numero delle variabili: un problema complesso è caratterizzato da *n* variabili e analizzare la distribuzione di probabilità congiunta completa è molto difficile quando il loro numero è elevato. Inoltre, spesso capita che non esistano nemmeno dati sufficienti per valutare ogni singola combinazione di eventi. In questi casi, il vantaggio derivante dall'uso di questo strumento è ulteriore.
L'utilizzo di una rete bayesiana, tuttavia, non implica che sia bayesiano anche il metodo con il quale sono stimati i parametri che spiegano le relazioni tra variabili. Data la struttura, o meglio il grafo, questi possono essere stimati anche sulla base dell'approccio inferenziale frequentista.
Una rete bayesiana si configura come un grafo costituito da nodi, che denotano variabili causali, e da archi orientati, l'orientamento dei quali sta ad indicare una relazione tra nodo genitore e nodo figlio, che, nel caso più informativo, possono rappresentare relazioni di causa effetto[8] e che, più in generale, indicano, invece, la mancanza di indipendenza statistica.

Altra peculiarità di questi grafi è quella di essere non ciclici, prevedendo, perciò, percorsi che non visitano più di una volta lo stesso nodo. L'acronimo utilizzato per definire questo tipo di strutture è DAG, dall'inglese "direct acyclic graph".

Le tre strutture locali tipiche che compongono i DAG bayesiani sono:

- la catena

**Figura 2.1**

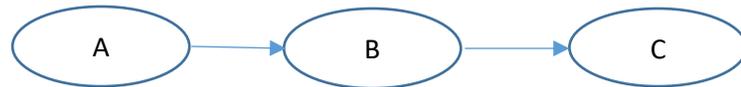

- la causa comune

**Figura 2.2**

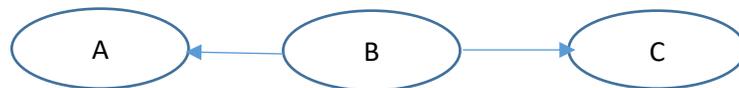

- l'effetto comune (collider)

**Figura 2.3**

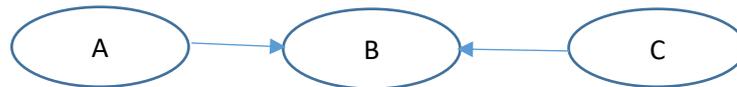

Ognuna di queste strutture rappresenta, in maniera molto intuitiva, il tipo di relazioni che intercorrono tra triplette di variabili.

Nonostante lo scheletro (ossia la struttura tra nodi ed archi, immaginando che questi ultimi non siano orientati) delle relazioni tra *A*, *B* e *C* sia identico:

- la catena (Fig. 2.1) e la causa comune (Fig. 2.2) implicano un'indipendenza tra i nodi estremi (*A* e *C*) condizionata al nodo intermedio (*B*), che nel primo caso funge da tramite, nel secondo è causa comune di due nodi discendenti;
- il collider (Fig. 2.3) implica che il condizionamento rispetto al nodo intermedio unisca (o d-connetta) i nodi estremi, tra loro indipendenti marginalmente.

Nei primi due casi, dunque, il nodo intermedio rende indipendenti due variabili che, in assenza di questo, verrebbero considerate associate; nel terzo caso, invece, è il nodo intermedio che rende dipendenti due variabili che, altrimenti, sarebbero considerate indipendenti. Due DAG che condividono lo stesso scheletro e si differenziano soltanto per l'orientamento di catene e strutture-causa comune, sono equivalenti. Allorquando vi fossero differenze relative ad un collider (per esempio, il grafo *G* presenta una catena tra *A*, *B* e *C* ed il grafo *G'*, con scheletro identico, presenta un collider con *B* quale effetto comune di *A* e *C*), i due grafi non sono, invece, equivalenti.

---

[8] L'interpretazione causale presuppone che, se si intervenisse sulla variabile genitore forzandone il valore (attraverso la funzione generalmente indicata come $do(X = x)$ ), avremmo una conseguente modifica della distribuzione riferita al nodo figlio. Cosa, questa, che presuppone la possibilità di sperimentare che ciò sia vero (Pearl, 1995). Per questo, particolare attenzione merita l'interpretazione, in termini causali, degli archi orientati tra nodi quando la struttura della rete è frutto di apprendimento automatico (machine learning). Salvo l'esistenza di talune condizioni che rendono la cosa possibile anche ad algoritmi come l'Inductive Causation (IC), è l'uomo a poter definire un rapporto di causazione, essendo la macchina in grado di rilevare per lo più relazioni di tipo associativo.

Quanto appena detto rende particolarmente importante la realizzazione dei test di d-separazione. I test di d-separazione hanno lo scopo di individuare, dati due nodi presenti nella rete *X* e *Y*, il set di nodi *Z* che li d-separa. Questi saranno rappresentati da tutti quei nodi che si posizionano lungo i path, diretti e indiretti, che intercorrono tra *X* e *Y*. Fanno, perciò, parte del set *Z*:

- i nodi che si posizionano lungo path-catene tra *X* e *Y* (rendono *X* e *Y* condizionatamente indipendenti);
- i nodi che si posizionano lungo path che uniscono *X* e *Y* a genitori comuni (rendono *X* e *Y* condizionatamente indipendenti);
- i nodi che si posizionano come collider lungo i path tra *X* e *Y* (non rendono *X* e *Y* condizionatamente indipendenti)[9].

Grazie all'uso dei test di d-separazione è possibile individuare, per ogni nodo, il Makov blancket, ossia il set di nodi che d-separa il nodo esaminato da tutti gli altri. In linea generale, esso sarà composto da: i nodi genitori; i nodi figli; i nodi sposi, vale a dire i nodi marginalmente indipendenti da quello in esame, che con esso condividano figli.

Una proprietà che distingue le reti bayesiane da un qualsiasi altro grafo aciclico è quella di Markov:

$$X_v \perp\!\!\!\perp X_{V \setminus \text{de}(v)} \mid X_{\text{pa}(v)} \quad \text{for all } v \in V \qquad (2.6)$$

Essa afferma che, a livello locale, i nodi $X_v$ sono d-separati dal set di nodi non appartenenti al set dei discendenti $X_{V \setminus de(v)}$ per il tramite dei nodi genitori $X_{pa(v)}$. Generalizzando, questa proprietà afferma l'indipendenza tra futuro e passato condizionatamente al presente e quindi, nel caso di una rete, tra i nodi precedenti rispetto ai successivi per il tramite di quelli intermedi.

Fondamentale, all'interno dei DAG, è il ruolo dei nodi. Ogni nodo si caratterizza per una distribuzione di probabilità riferita alle diverse modalità che la variabile rappresentata può assumere. Essa è stimata attraverso i dati osservati, oppure è elicitata dall'esperto. Tra due nodi consecutivi vi è una tabella di probabilità condizionate (o CPT, dall'acronimo inglese "conditional probabilities table"): essa rappresenta le probabilità con cui le diverse modalità del nodo figlio si presentano condizionatamente alle modalità che può assumere il nodo genitore. Per i nodi-radice[10], quelli che si situano ai vertici iniziali della rete, la CPT corrisponde alle probabilità delle diverse modalità che la variabile stessa assume.

Spesso, quando la rete risultasse troppo complessa o in presenza di strutture locali cicliche (Fig. 2.4), utile potrebbe rivelarsi la soluzione di unire, fondere, in un'unica variabile/nodo due o più variabili tra loro associate. Per esempio, immaginando di avere due variabili boleane (ognuna delle quali presenta due modalità), si potrebbe procedere sostituendole con un'unica variabile che presenti quattro modalità, ognuna rappresentativa delle quattro combinazioni possibili: vero/vero; vero/falso; falso/vero; falso/falso. Ciò è utile non soltanto per alleggerire la rete da nodi superflui, ma anche per evitare, lì dove l'apprendimento fosse realizzato automaticamente, che due grafi equivalenti possano non risultare tali. La struttura in Fig. 2.4 può essere vista come costituita da due catene (*A->B->D, A->C->D*), oppure da una forchetta ed un collider (*B<-A->C, B->D<-C*). Dal momento che, per la macchina, l'orientamento che assume una catena è poco rilevante e viene determinato in modo da evitare che il DAG diventi ciclico, può accadere che, diversamente da quanto presente in figura, gli archi puntino da *D* verso *A* per il tramite di *B* e *C*. Ciò che a livello locale non produce effetti (l'orientamento delle due catene), a livello globale sì poiché, come detto in precedenza, oltre allo scheletro, è il posizionamento dei collider ad essere importante nella misurazione dell'equivalenza tra due reti. Nell'esempio, tale equivalenza verrebbe meno (il collider non sarebbe più *B->D<-C*, bensì *B->A<-C*).

---

[9] Nella prima definizione di collider erano solo considerati i nodi figli di genitori marginalmente indipendenti. Grazie a Pearl (1988) saranno considerati tali anche i discendenti di questi nodi figli.

[10] I nodi-radice sono anche classificati come variabili esogene, non avendo, queste, nessun ascendente all'interno della rete ma almeno un discendente. Alternative a queste sono le variabili endogene, quelle che, cioè, hanno almeno un ascendente all'interno della rete, pur non dovendo necessariamente avere dei discendenti.

**Fig. 2.4**

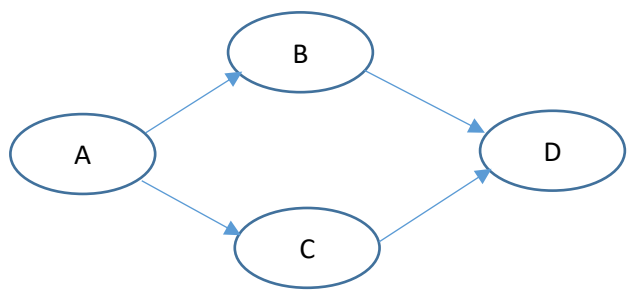

## 3. UNA RETE BAYESIANA SUI DATI EDIT 2015

Per quanto concerne la parte applicativa, lo strumento utilizzato in fase preparatoria dei dati e per la loro analisi successiva è R. In particolare, si è fatto uso dei pacchetti `bnlearn` (Scutari e Denis, 2014), `gRain` (Højsgaard, 2012), `Rgraphviz` (Hansen et al., 2016), `ggm` (Marchetti et al., 2015).

### 3.1 La preparazione dei dati e l'apprendimento della rete

Il database iniziale su cui si è lavorato contiene circa 160 variabili (il database completo dell'Edit è più grande, all'incirca 200 variabili) per 5077 osservazioni. Una prima operazione di "pulitura" riguarda la selezione delle variabili ritenute più importanti per gli scopi dell'analisi, arrivando così ad averne circa una sessantina.
Dopo attenta analisi delle variabili selezionate, in alcuni casi si è proceduto con una riclassificazione, in modo tale da poter ridurre i livelli di alcune variabili discrete (per esempio, ore di sonno, giudizio relativo ai rapporti con coetanei e genitori, giudizio relativo al rendimento scolastico, esito del test K6, etc.), o rendere discrete delle variabili continue (per esempio, l'indice di massa corporea).
Per variabili fortemente associate sulla base di elementi non osservati nell'indagine e di complessa quantificazione nella realtà, si è proceduto con la creazione di variabili fittizie frutto dell'unione tra variabili originarie. È questo il caso di tutte le variabili riferite ai genitori singolarmente presi. Per esempio, piuttosto che avere informazioni separate sulla classe d'età del padre e della madre, dopo aver ridotto il numero di modalità (sotto i 50 anni, sopra i 50 anni, non so), si è creata una variabile unica caratterizzata da livelli riguardanti la concordanza di età: entrambi sotto i 50 anni; entrambi sopra i 50 anni; uno sopra e l'altro sotto i 50; non so. Procedimento simile si è seguito per titolo di studio, condizione professionale e abitudine all'assunzione di alcolici ed al fumo, sempre riferiti ai genitori dell'intervistato singolarmente presi.
Dal momento che l'algoritmo utilizzato non ammette valori NA, si è poi intervenuto su alcune risposte sostituendo tali valori. Ci si riferisce al caso delle domande a cascata: nel momento in cui l'intervistato risponde di no alla prima domanda, quelle successive non gli vengono poste. Per esempio, se al soggetto viene chiesto se ha assunto cannabis nell'ultimo anno e la risposta è negativa, non gli viene domandato né se ha assunto cannabis nell'ultimo mese, né, tantomeno, la frequenza di assunzione nell'ultimo mese. In casi come questo si è proceduto sostituendo il valore NA, sulla base delle risposte date alle domande precedenti. In questa maniera si è evitato di perdere troppe osservazioni a causa di valori mancanti.
Il database che, a seguito delle diverse fasi di filtraggio, è usato nell'apprendimento automatico contiene 3.647 osservazioni e 48 variabili.
Ultimo accorgimento è quello di sfruttare l'argomento "blacklist" dell'algoritmo Hill-Climbing utilizzato per inferire reti bayesiane dai dati. Questa opzione consente di specificare le variabili tra le quali non si vuole che si posizioni un arco orientato. L'opzione è particolarmente utile quando si vuol provare a dare un connotato di causalità ad una rete. Per questo motivo, si sono imposti dei limiti all'apprendimento automatico finalizzati a rendere nodi-radice, dunque variabili esogene al modello: il sesso dell'intervistato; l'età dell'intervistato; la concordanza di età tra i genitori. Per la variabile riferita alla concordanza di titolo di studio tra i genitori, si è lasciato all'algoritmo la scelta relativa al rendere questa variabile esogena o, al massimo, figlia della sola variabile riferita alla concordanza d'età tra genitori.

Di seguito è riportato l'elenco completo delle variabili del database finale, con relativo codice utile ad individuare la posizione dei nodi ad esse associati all'interno dei grafi presentati più avanti:
- A Sesso
- B Età
- C Provincia di residenza
- D Comune di residenza
- M Genitori separati
- N Classe qualità rapporti familiari
- O Classe qualità rapporti coi coetanei
- P Classe rendimento scolastico
- Q Ripetizione di anni scolastici
- R Uso del pc
- S Ore giornaliere al pc
- Tt Classe numero libri letti nell'anno
- U Classe ore di sonno per notte
- V Propensione a nascondere cifre scommesse
- W Impulso a scommettere cifre sempre più alte
- X Classe quantità sigarette fumate al giorno
- Y Episodi ubriachezza nell'ultimo anno
- Z Abuso di alcol nell'ultimo mese
- AE Uso di cannabis nell'ultimo anno
- AF Uso altre droghe nell'ultimo anno
- AG Uso cannabis nell'ultimo mese
- AH Uso altre droghe nell'ultimo mese
- AI Frequenza uso cannabis nell'ultimo mese
- AJ Frequenza uso altre droghe nell'ultimo mese
- AK Classe multe subite per infrazioni codice stradale
- AL Bullismo subito
- AM Bullismo fatto
- AN Aver seguito una dieta nell'ultimo mese
- AO Aver fatto attività sportiva per perdere peso nell'ultimo mese
- AP Aver vomitato per perdere peso nell'ultimo mese
- AQ Aver Assunto di pillole per perdere peso nell'ultimo mese
- AR Aver fatto regolarmente sport nell'ultimo anno
- AS Classe n. volte guida sotto effetti stupefacenti
- AT Classe n. incidenti con ricorso a pronto soccorso
- AU Classe propensione uso droghe da solo o in compagnia
- AV Classe età prima esperienza droghe
- AW Uso di alcol prima ultimo rapporto sessuale
- AX Uso profilattico ultimo rapporto sessuale
- AY Classe metodi anticoncezionali ultimo rapporto
- AZ Uso della pillola del giorno dopo

- BA Tipo di scuola
- BB Classe score k6 sul distress
- BC Classe indice massa corporea
- BD Concordanza età dei genitori
- BE Concordanza titolo di studio dei genitori
- BF Concordanza condizione professionale dei genitori
- BG Concordanza utilizzo alcolici da parte dei genitori
- BH Concordanza abitudine al fumo dei due genitori

**3.2 L'algoritmo Hill-Climbing e i criteri di regolarizzazione**

L'algoritmo utilizzato per la costruzione di reti bayesiane sulle variabili della banca dati Edit 2015 è l'Hill-Climbing (HC), letteralmente "scalata della collina", espressione che descrive bene il comportamento dell'algoritmo: lo scopo è di massimizzare una funzione obbiettivo, definibile come un'utilità, e raggiungere l'ottimo, ossia il massimo di utilità (da ciò l'idea di scalare). Si può anche cambiare il segno, trasformando l'utilità in una funzione di perdita da minimizzare e, in maniera figurativa, la scalata di una collina in una discesa in un pozzo. Come si può evincere dall'esempio in R, riportato nel seguito, il metodo di funzionamento è piuttosto semplice.

**Esempio in R**[11]
```
#' @title Hill climbing
#'
#' @name hillclimbing
#' @rdname hillclimbing
#'
#' @description
#' Use hill climbing to find the global minimum
#'
#' @param f function representing the derivative of \code{f}
#' @param x an initial estimate of the minimum
#' @param h the step size
#' @param m the maximum number of iterations
#'
#' @details
#'
#' Hill climbing
#'
#' @return the \code{x} value of the minimum found
#'
#' @family optimz
#'
#' @examples
 f <- function(x) {
    (x[1]^2 + x[2] - 11)^2 + (x[1] + x[2]^2 - 7)^2
 }

#' @importFrom stats runif
#' @importFrom stats rnorm

#' @export

 hillclimbing <- function(f, x, h = 3, m = 10000) {
  n <- length(x)

  xcurr <- x
  ycurr <- f(x)

  for(i in 1:m) {
     xnext <- xcurr
     i <- ceiling(runif(1, 0, n))
     xnext[i] <- rnorm(1, xcurr[i], h)
     ynext <- f(xnext)
```

---
[11] Fonte: https://github.com/cran/cmna/blob/master/R/hillclimbing.R

```
   if(ynext < ycurr) {
      xcurr <- xnext
      ycurr <- ynext
   }
 }

 return(xcurr)
}

hillclimbing(f, c(0,0))
hillclimbing(f, c(-1,-1))
hillclimbing(f, c(10,10))
```

Considerando un altro esempio, si immagini di avere uno scheletro costituito da tre nodi *A*, *B* e *C*, e due archi non orientati. L'algoritmo deve decidere se sia conveniente, in termini di massimizzazione della funzione obbiettivo, aggiungere altri nodi allo stesso oppure no (Fig. 3.1). Peculiarità dell'HC è la procedura per passi. Viene valutato se l'aggiunta del nodo *D* produce un miglioramento sotto il profilo della massimizzazione della funzione obbiettivo: se la risposta è affermativa, il nodo viene incorporato nella struttura e si procede provando ad incorporare anche il nodo rimanente; se la risposta è negativa, la struttura rimane così com'è e l'algoritmo valuta cosa accade se alla struttura si aggiunge il nodo *E*.

**Fig. 3.1**

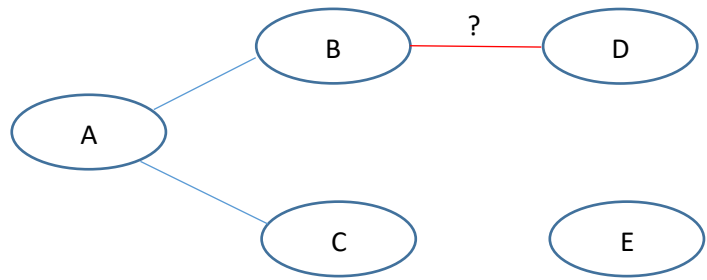

La funzione obbiettivo è generalmente costituita da una funzione di perdita (o loss) e da un elemento di regolarizzazione il cui fine è quello di penalizzare i modelli troppo complessi, in modo da scongiurare overfitting.

Per quanto concerne la funzione di perdita $l_x(\hat{\theta}, \theta^*)$, obbiettivo è quello di individuare i parametri $\hat{\theta}$ che meglio descrivono i dati *x* su cui è costruita la rete, sapendo che il valore reale di questi è $\theta^*$.

$$L_x(\hat{\theta}) = \langle l_x(\hat{\theta}, \theta^*) \rangle_{P(\theta=\theta^*|x)} = \int l_x(\hat{\theta}, \theta^*) * P(\theta = \theta^*|x) * d\theta^* \quad (3.1)$$

Il valore atteso della funzione di loss è uguale alla media ponderata di tutti i possibili valori $\theta$, il peso dei quali è definito dalle probabilità relative che questi valori hanno di essere presi in considerazione. Il valore ottimale di $\hat{\theta}$ è quello che maggiormente si avvicina a $\theta^*$, con conseguente minimizzazione delle perdite.

Per quel che concerne le modalità per calcolare la penalità, a seconda del metodo utilizzato, la struttura della rete è destinata a cambiare. Di seguito sono brevemente descritti tre tra i principali criteri.

Il Bayesian Information Criterion (BIC) è uguale a:

$$BIC = \ln(n)k - 2\ln(L) \quad (3.2)$$

Dove *L* indica la funzione loss, *k* Il numero dei parametri e *n* il numero di osservazioni.
L'Akaike Information Criterion (AIC) prevede:

$$AIC = 2k - 2\ln(L) \quad (3.3)$$

Pur essendo molto simile alla (3.2), la penalità è meno consistente, con possibili effetti sull'eccessivo overfitting del modello.

Il Bayesian Dirichlet (BD), invece, adotta una strategia diversa (Heckerman et al., 1995). Mentre con i due criteri visti finora viene cercata la rete che, singolarmente presa, raggiunge il miglior equilibrio tra funzione loss e penalità, il BD consente di individuare la classe di reti che meglio soddisfa i due obbiettivi. Lo score espresso in questi termini, dunque, non si riferisce ad una rete in particolare, ma ad un insieme di queste che condividono stesso scheletro e stesse strutture a V, o collider. Lo score, dunque, valuta le diverse classi di reti rispetto ad un'ipotetica classe a-priori elicitata dall'esperto. Individuando con $N'$ le istanze riferite alla classe ipotetica di reti a-priori e con $N$ quelle che, invece, si riferiscono a quanto osservato, il BD score, nella sua versione generale, si presenta come segue:

$$BD = \ln(L) + \prod_{i=1}^{N} \prod_{j=1}^{q_i} \left( \ln\left(\frac{\Gamma(N'_{ij})}{\Gamma(N_{ij} + N'_{ij})}\right) + \prod_{k=1}^{r_i} \ln \frac{\Gamma(N_{ijk} + N'_{ijk})}{\Gamma(N'_{ijk})} \right) \quad (3.4)$$

$N_{ijk}$ rappresenta il numero delle istanze riferite alle osservazioni i-esime che presentano la modalità k-esima riferita alla variabile/nodo figlio in corrispondenza della j-esima variabile/nodo genitore. È, dunque, ciò che determina il parametro che congiunge la distribuzione della variabile genitore con quella del figlio. $N_{ij}$, invece, rappresenta le istanze riferite alle modalità della sola variabile/nodo genitore, e determina, dunque, il parametro che spiega la distribuzione di questa.

Dal momento che l'elicitazione della a-priori nelle reti è qualcosa di difficile da fare, soprattutto in presenza di un gran numero di variabili, il likelihood equivalent Bayesian Dirichlet (BDe) risolve la questione lasciando allo studioso la scelta relativa alla determinazione dell'iperparametro $N'$, sapendo che:

$$N'_{ijk} = N' * P(X_i = x_i, \Pi_i = w_{ij} | G) \quad (5)$$

$$N'_{ij} = N' * P(\Pi_i = w_{ij} | G) \quad (6)$$

Con $G$ che rappresenta lo scheletro della classe di reti considerata, $\Pi_i$ che individua il set delle variabili/nodo genitore e $w_{ij}$ il valore specifico che queste variabili assumono rispetto alle figlie.

L'iperparametro si configura, perciò, come una sorta di insieme di osservazioni sulla base delle quali si è formata la conoscenza che lo studioso ha, a-priori, su quell'ambito. Nello specifico, però, assume anche il ruolo di peso attribuito alla regolarizzazione nella determinazione dello score finale: se $N'$ è settato ad un valore elevato, la classe di reti privilegiata sarà poco parsimoniosa, viceversa quando $N'$ assume un valore piccolo. È anche questa possibilità di determinare il peso della penalità esternamente che contraddistingue il BDe dai criteri BIC e AIC.

## 4. RISULTATI

Il primo passo è stato quello di creare alcune reti bayesiane, ognuna delle quali caratterizzata dall'utilizzo di un metodo di regolarizzazione differente. In tal modo, valutando più alternative, è stato possibile scegliere quella che meglio concilia capacità predittiva e parsimonia.

Le variabili sono state codificate come riportato nel paragrafo 3.1.

Il risultato ottenuto utilizzando la penalità BIC è:

```
model:
  [A][B][D][R][BD][BE][Q|B][S|R][AN|A][AX|B][BC|A][BF|BD][P|Q][AO|A:AN][AQ|AN][AY|AX][BA|Q:BE]
```

```
[C|BA][Tt|A:BA][AP|A:AQ][AR|A:AO][AW|AY][BB|A:P][BG|BA][O|BB][U|BB][AF|AW][AZ|AW][BH|BG][M|BH]
[AH|AF][N|M:BB][AJ|AH][AU|AH:AY][AV|AU][AE|AF:AV][AM|AV][Y|AE:AW][AG|AE][AL|AM:BB][V|A:Y][X|AG]
[Z|Y:AG][AI|AG][W|V][AS|AI][AK|AS][AT|AS]
nodes:                                    48
arcs:                                     55
  undirected arcs:                        0
  directed arcs:                          55
average markov blanket size:              2.75
average neighbourhood size:               2.29
average branching factor:                 1.15

learning algorithm:                       Hill-Climbing
score:                                    BIC (disc.)
penalization coefficient:                 4.10083
tests used in the learning procedure:     3708
optimized:                                TRUE
```

La prima parte dell'output, ossia la descrizione della struttura, è più semplice da valutare ricorrendo alla rappresentazione grafica tramite DAG (pag. 19). Tra le sintesi più importanti c'è il numero degli archi orientati e dei nodi, nel caso specifico rispettivamente 55 e 48, che costituiscono i parametri del modello. In relazione a quanto affermato con riferimento al concetto di d-separazione, è importante anche la dimensione media del Markov blanket, pari, in questo caso, a 2,75 per nodo.

Si consideri, ora, ciò che si ottiene, a parità di condizioni di partenza (ci si riferisce alle variabili utilizzate ed alle blacklist), utilizzando penalità AIC:

```
model:
 [A][B][BD][AX|A:B][BE|BD][C|BE][AY|AX][BF|BD:BE][D|C][AW|A:AY][BG|BF][AF|A:AW][AZ|A:AW][BA|D]
 [Q|B:BA][R|BA][Tt|A:BA][AH|AF][S|R][AJ|AH][AU|AJ:AW][AV|AH:AU][AE|AH:AV][AG|AE:AH][AI|AG]
 [X|AI:AY][Y|AI:AW][AS|A:AI][P|A:Q:X][Z|Y:AG][AK|AH:AS][AN|A:Y:AW][AT|AS:AZ][BH|X:BG][M|BG:BH]
 [AQ|AH:AN][BB|A:P:Y][N|M:BB][U|AV:BB][AP|A:AE:AQ][O|N:BB][AO|A:AN:AP][AL|O:BB][AR|A:AO:BA]
 [BC|A:AO][AM|A:AL:AV][V|A:Y:AM][W|V]
nodes:                                    48
arcs:                                     86
  undirected arcs:                        0
  directed arcs:                          86
average markov blanket size:              4.83
average neighbourhood size:               3.58
average branching factor:                 1.79

learning algorithm:                       Hill-Climbing
score:                                    AIC (disc.)
penalization coefficient:                 1
tests used in the learning procedure:     5119
optimized:                                TRUE
```

Questa seconda rete è più complessa della prima. Ciò lo si evince dal fatto che è maggiore il numero di archi orientati. Dalla maggior complessità discende che ogni nodo ha, in media, una dimensione del Markov blancket quasi doppia rispetto alla rete ottenuto con BIC.

Infine è generata una rete adoperando penalità BDe con iperparametro, riferito al campione immaginario (o equivalente), pari a 10:

```
model:
 [A][B][C][BD][D|C][AX|B][BE|BD][BF|BD][AY|AX][BA|D][BG|BF][Q|B:BA][Tt|A:BA][AR|A:BA][AW|AX:AY]
 [AF|AW][AZ|A:AW][AH|AF][AJ|AF:AH][AU|AJ:AW][AV|AH:AU][AE|AH:AV][Y|AE:AW][AG|AE:AF:AU][V|A:Y]
 [Z|Y:AG][AI|AE:AF:AG][AN|A:Y][W|V][X|AI:AY][AO|A:AN:AR][AQ|AH:AN][AS|A:AI][P|A:Q:X][AP|A:AE:AQ]
 [BC|A:AO][BH|X:BG][M|BH][BB|A:P][N|M:BB][U|X:BB][R|N][O|R:BB][S|R][AL|O:BB][AM|A:AL:AV][AK|AM:AS
] [AT|AK:AS]
nodes:                                    48
```

```
arcs:                                    81
  undirected arcs:                        0
  directed arcs:                         81
average markov blanket size:             4.58
average neighbourhood size:              3.38
average branching factor:                1.69

learning algorithm:                      Hill-Climbing
score:                                   Bayesian Dirichlet (BDe)
graph prior:                             Uniform
imaginary sample size:                   10
tests used in the learning procedure:    4978
optimized:                               TRUE
```

Questa terza opzione si posiziona tra le altre due, avendo 81 archi orientati ed una dimensione media del Markov blanket poco più bassa rispetto alla rete con penalità AIC. Questo è ciò che risulta utilizzando l'iperparametro di default, pari a 10, che rappresenta il valore ritenuto migliore. Gli effetti prodotti dall'aumentare e ridurre dell'iperparametro di un ordine di grandezza, portandolo a 100 e ad 1, sono: nel primo caso, una rete con 63 archi orientati e una dimensione media del Markov blanket pari a 3,29; nel secondo caso, una rete con 130 archi orientati ed una dimensione media del Markov blanket pari a 8,08 (risultati non mostrati).

Le informazioni ricavate nei tre tentativi possono essere riassunte dalla rappresentazione grafica delle reti, ottenuta tramite DAG. L'impatto visivo di questi grafi è tale da dare un'immediata idea della parsimonia di un modello, nonché delle relazioni tra nodi/variabili.

**Figura 4.1** Rete bayesiana con penalità BIC. La corrispondenza tra nodi e variabili è riportata nel paragrafo 3.1

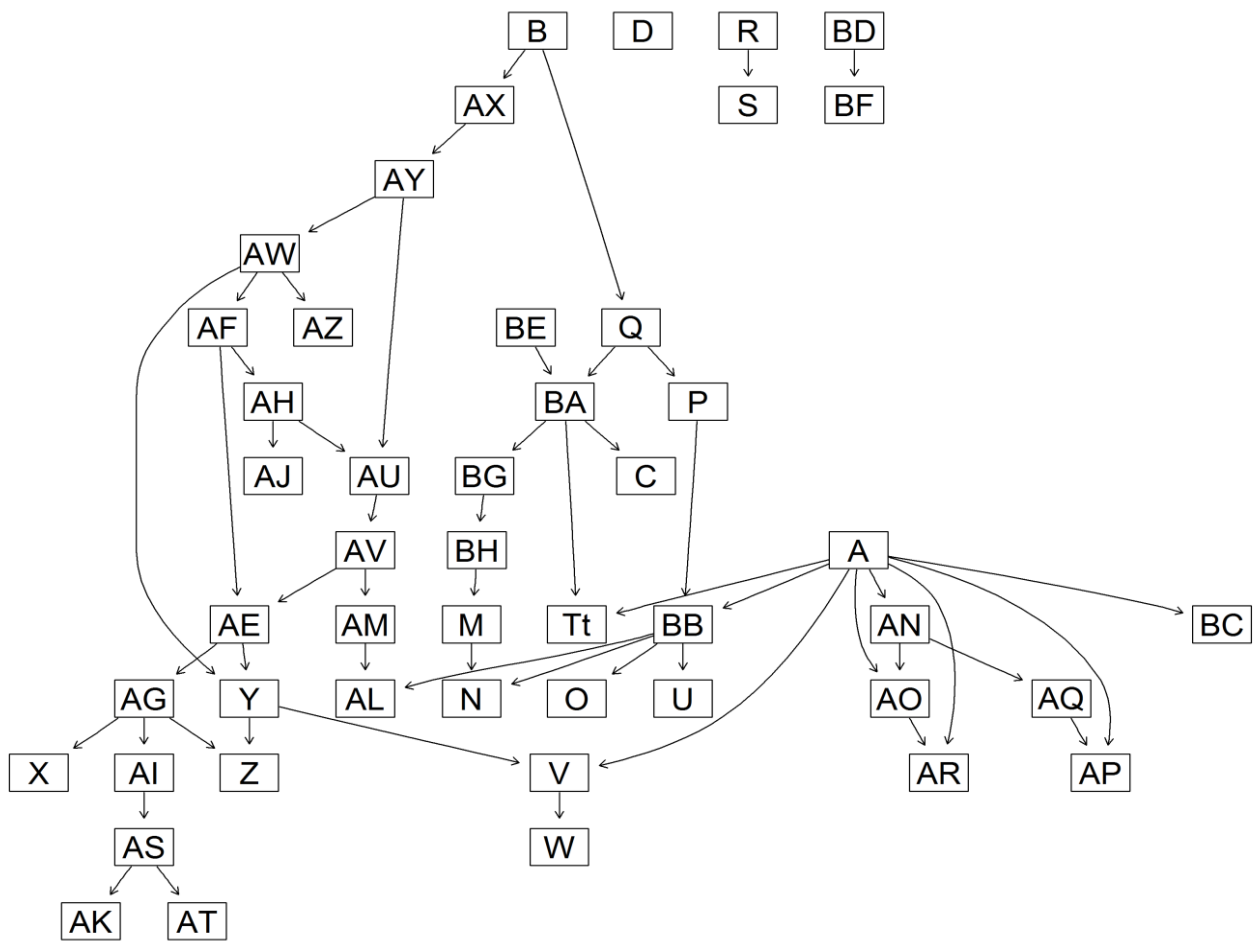

**Figura 4.2** Rete bayesiana con penalità AIC. La corrispondenza tra nodi e variabili è riportata nel paragrafo 3.1

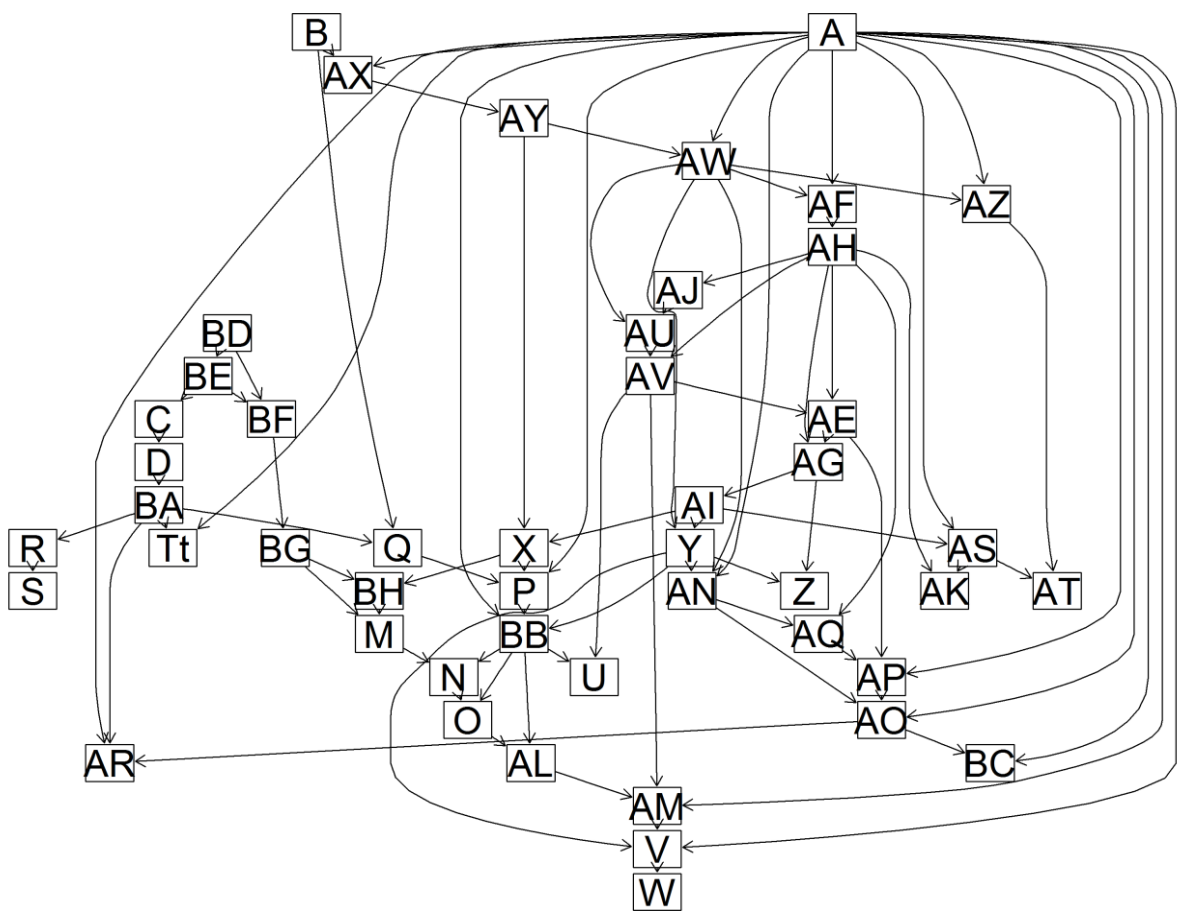

**Figura 4.3** Rete bayesiana con penalità BDe. La corrispondenza tra nodi e variabili è riportata nel paragrafo 3.1

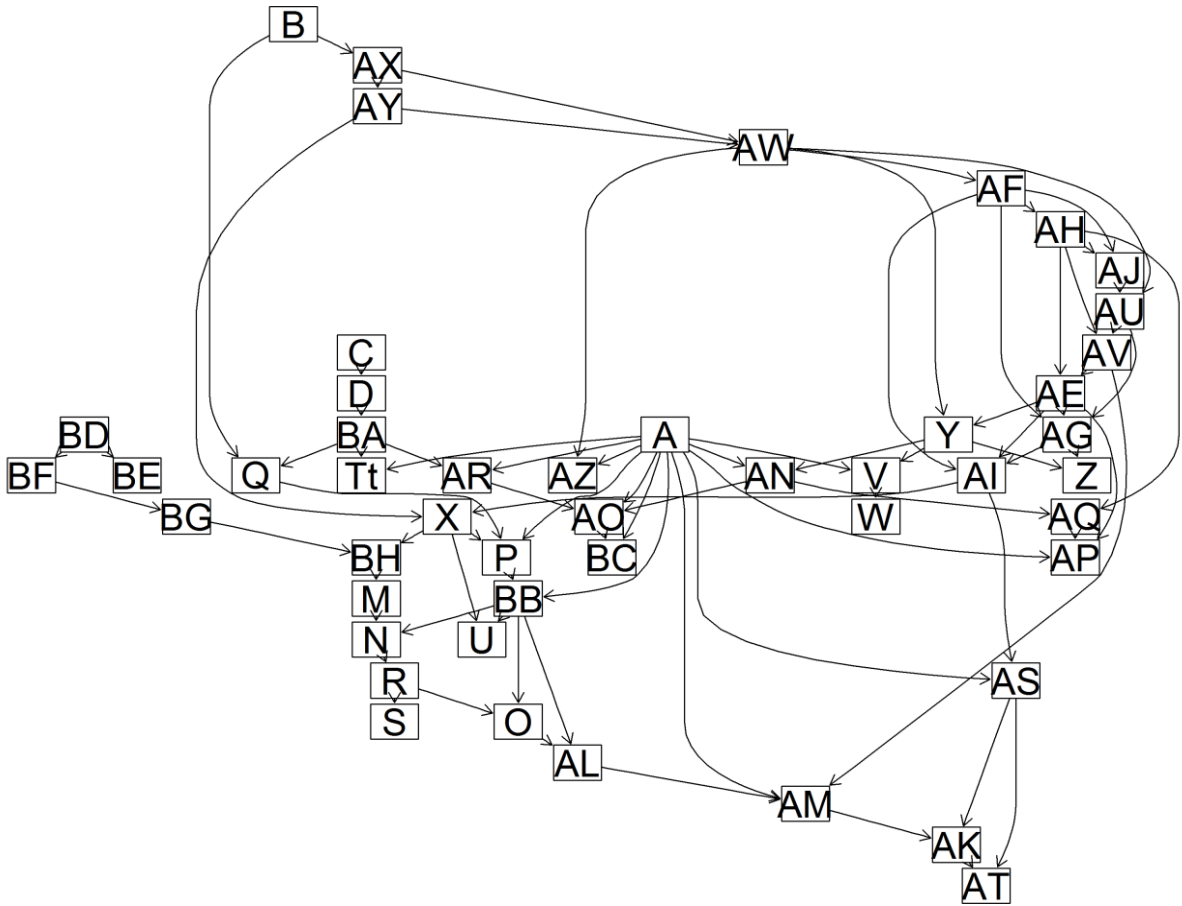

Premesso che la maggior parte delle relazioni si ripresenta nella stessa maniera in tutti e tre i modelli, si può ritenere che la rete migliore sia quella con penalità BDe (Fig. 4.3). Questo criterio, infatti, assegna stesso score a reti che sono equivalenti per valore di verosimiglianza. In aggiunta a ciò, la rete con criterio BDe, da un lato non esclude nessun nodo dal DAG principale, come avviene con la rete con penalità BIC (Fig. 4.1), dall'altro lo fa con un numero di parametri contenuto, a differenza della rete generata con penalità AIC (Fig. 4.2).
Sulla base delle relazioni definite dal modello prescelto, genitori e figli dei nodi più direttamente legati al tema dei comportamenti rischiosi alla guida sono:
• per AS – numero di volte, raggruppate in classi, in cui l'intervistato ha guidato sotto effetto di stupefacenti, i nodi genitori sono:
✓ A – genere;
✓ AI – frequenza dell'uso di cannabis nell'ultimo mese;
i nodi figli sono:
✓ AK – numero di multe subite per infrazioni al codice della strada, raggruppate in classi;
✓ AT – numero di volte, raggruppate in classi, in cui si è subito un incidente stradale tale da richiedere ricovero al pronto soccorso;
• per AK – numero di multe subite dall'intervistato per infrazioni al codice della strada, raggruppate in classi, i nodi genitori sono:
✓ AM – bullismo perpetrato verso gli altri;
✓ AS – numero di volte, raggruppate in classi, in cui l'intervistato ha guidato sotto effetto di stupefacenti, il nodo figlio è:

✓ AT – numero di volte, raggruppate in classi, in cui si è subito un incidente stradale tale da richiedere ricovero al pronto soccorso;
- per AT – numero di volte, raggruppate in classi, in cui l'intervistato ha subito un incidente stradale tale da richiedere ricovero al pronto soccorso, i nodi genitori sono:

✓ AK – numero di multe subite per infrazioni al codice della strada, raggruppate in classi;

✓ AS – numero di volte, raggruppate in classi, in cui l'intervistato ha guidato sotto effetto di stupefacenti, i nodi figli non sono presenti essendo, questo, un "nodo-foglia".

Quanto appena descritto è utile per individuare quali variabili sarebbe sufficiente osservare se si volessero prevedere i comportamenti rischiosi degli adolescenti alla guida, rendendo più parsimonioso il questionario. Nello specifico, basterebbe conoscere il genere degli adolescenti, la frequenza dell'uso di cannabis nell'ultimo mese e l'eventuale bullismo esercitato verso coetanei per avere informazioni sufficienti a ricavare una distribuzione di probabilità per ognuno dei comportamenti di guida considerati.

È opportuno ri-sottolineare che l'arco orientato che l'apprendimento automatico rileva tra due nodi non implica necessariamente un nesso di causalità. Per cui, dire, per esempio, che conoscere il genere di un adolescente permette di stimare il numero di volte in cui egli guiderà sotto effetto di stupefacenti è altro rispetto ad affermare che il genere "causa" la maggior o minor propensione alla guida sotto effetti di stupefacenti. Da ciò l'esigenza di sottoporre le reti bayesiane frutto di apprendimento automatico a validazione da parte di esperti che, eventualmente, battezzeranno la rete come causale, qualora vi siano le condizioni per farlo. Allorquando una relazione tra due nodi sia meramente associativa, la rete esaminata da un esperto può suggerire la presenza di variabili latenti, ossia variabili non osservate che potrebbero inserirsi "a metà" di un arco orientato, oppure essere genitore o figlio comune, arricchendo la banca dati di informazioni utili per trasformare una rete probabilistica in causale.

Interessante è, a questo punto, applicare i test di d-separazione per capire se la conoscenza dei comportamenti rischiosi alla guida è sufficiente per risalire a conoscere un'altra informazione importante che viene raccolta all'interno della banca dati Edit, vale a dire il punteggio relativo al Kessler Psychological Distress Scale (o K6). Esso è un indice ricavato dalle risposte ad alcune domande sottoposte agli intervistati relativamente alla frequenza di stati d'animo "a rischio" (sentirsi nervosi, agitati, depressi, inutili, etc.) nell'ultimo mese. Si tratta di un'informazione che, non soltanto riassume la condizione psicologica dei ragazzi, ma segnala anche una maggior o minor probabilità, per gli stessi, di transitare in una condizione di malattia psichiatrica. Il nodo riferito al punteggio di questo test all'interno della rete con penalità BDe è quello con sigla BB. Suoi nodi genitori sono:
- A – genere;
- P – livello del rendimento scolastico.

Suoi nodi figli sono:
- N – qualità dei rapporti familiari;
- U – numero di ore di sonno per notte, raggruppate in classi;
- O – qualità dei rapporti con i coetanei;
- AL – bullismo subito.

L'idea alla base dei test di d-separazione che seguono è: conoscendo i comportamenti rischiosi tenuti da un adolescente alla guida, è possibile risalire al valore dell'indice di distress dello stesso conoscendo anche altre informazioni quali uso e abuso di alcolici, uso e abuso di droghe, abitudini alimentari patologiche, comportamenti sessuali a rischio, condizione socio economica? Per fare questo, sulla base di quanto affermato nel paragrafo 2.2, è sufficiente considerare il punteggio del K6 e le variabili relative a comportamenti rischiosi alla guida come nodi estremi di un set di path e vedere se le variabili che si riferiscono ai diversi comportamenti "condizionanti" su elencati si posizionano lungo questi diversi path, bloccandoli.

Di seguito i risultati dei test di d-separazione effettuati con il software R ed il package ggm:

**Tabella 4.1** Risultati test di d-separazione tra set di nodi più rilevanti

| Set nodi X | Set nodi Z | Set nodi Y | Conoscendo Y, posso risalire a X conoscendo anche Z? |
|---|---|---|---|
| Indice distress {BB} | Uso e abuso di droghe {AE, AF, AG, AH, AI, AJ, AU, AV} | Comportamenti rischiosi alla guida {AK, AS, AT} | FALSO |
| Indice distress {BB} | Uso e abuso di alcol {Y, Z} | Comportamenti rischiosi alla guida {AK, AS, AT} | FALSO |
| Indice distress {BB} | Condizioni socio economiche {M, BA, BE, BF} | Comportamenti rischiosi alla guida {AK, AS, AT} | FALSO |
| Indice distress {BB} | Comportamenti sessuali {AW, AX, AY, AZ} | Comportamenti rischiosi alla guida {AK, AS, AT} | FALSO |
| Indice distress {BB} | Comportamenti alimentari {AN, AO, AP, AQ, BC} | Comportamenti rischiosi alla guida {AK, AS, AT} | FALSO |

Come si può evincere dai risultati dei test, in nessun caso è possibile risalire al punteggio del test K6 conoscendo i comportamenti rischiosi alla guida, condizionatamente alla conoscenza di altre informazioni raggruppate per tipologia e riferite ad altri comportamenti a rischio o alle condizioni socio economiche.

Una volta definita una rete tra variabili, è possibile utilizzarla per fare inferenza che, nel caso specifico, può significare provare a prevedere i comportamenti a rischio alla guida. Per esempio, si immagini di voler prevedere la probabilità di incorrere in multe legate ad infrazioni stradali partendo dall'indice di distress di un adolescente. Innanzitutto, è utile capire quali sono i genitori del nodo BB, quello legato al distress, nonché le probabilità condizionali tra nodi genitori (in questo caso A - genere e P – rendimento scolastico) e nodo figlio:

```
Conditional probability table:

, , P = 0

   A
BB           1          2
  0 0.54506803 0.28282828
  1 0.42261905 0.55808081
  2 0.03231293 0.15909091

, , P = 1

   A
BB           1          2
  0 0.37711313 0.15929204
  1 0.53706112 0.54646018
  2 0.08582575 0.29424779

, , P = 2

   A
BB           1          2
  0 0.17777778 0.10000000
  1 0.53333333 0.40000000
  2 0.28888889 0.50000000
```

```
, , P = 99

    A
BB           1 2
  0 0.71428571
  1 0.28571429
  2 0.00000000
```

Sulla base dei dati Edit 2015, quando il nodo P assume modalità 0, vale a dire quando il rendimento scolastico è buono, per i maschi (A = 1) la probabilità di avere un indice di distress basso (BB = 0) è pari allo 0.54, mentre per le femmine (A = 2) è dello 0.28. Se, però, si considera il caso di studenti con rendimento basso (P = 2), la probabilità dei maschi di avere un indice di distress basso è pari allo 0.17 (quasi un terzo del caso precedente), mentre per le femmine è pari allo 0.1 (meno della metà del caso precedente). Di contro, se per chi ha rendimento scolastico buono, l'indice di distress è alto con probabilità 0.03 per i maschi e 0.16 per le femmine, per chi ha un rendimento scolastico negativo, queste probabilità salgono allo 0.28 per i maschi e allo 0.5 per le femmine.

Assodato come il nodo del distress è condizionato dai suoi genitori, è utile guardare alle distribuzioni marginali del nodo BB e del nodo AK, quello che vorremo stimare rispetto a dati futuri:

```
$BB
BB
        0         1         2
0.3705844 0.5071183 0.1222973

$AK
AK
         0          1         99
0.94089917 0.04651675 0.01258408
```

Per quel che concerne il distress (BB), la modalità più probabile è quella relativa al valore intermedio (0,51). Per quanto riguarda, invece, il numero di multe subite per comportamenti alla guida, raggruppate in classi (AK), la modalità più probabile è 0, ossia nessuna multa subita (0,94). Si evidenzia come la modalità "99", presente in questo, ma anche in altri casi, rappresenta sempre la risposta "non so/ non ricordo".

Ulteriori informazioni sono ricavabili guardando alla distribuzione congiunta tra le modalità dei due nodi in esame:

```
     AK
BB           0           1          99
  0 0.3487387 0.017400142 0.004445583
  1 0.4771005 0.023559509 0.006458252
  2 0.1150600 0.005557102 0.001680244
attr(,"class")
[1] "parray" "array"
```

Il caso più probabile è quello relativo ad adolescenti che presentano distress medio e nessuna multa subita per infrazioni al codice della strada (0,48). Tralasciando la modalità di chi non ricorda se ha mai subito multe (AK = 99), il caso meno probabile è quello di chi ha subito almeno una multa per infrazioni al codice della strada ed ha un indice di distress elevato (0,005).

La rete permette anche di valutare cosa succede tra BB e AK condizionando rispetto ad una terza variabile. Per esempio, si immagini di condizionare rispetto ad AL – bullismo subito, una delle variabili che si posiziona lungo il path tra i due nodi considerati. AL può assumere tre modalità: 1, per chi ha subito bullismo; 2, per chi non lo ha subito; 99, per chi non sa rispondere. Di seguito viene valutata la relazione tra BB e AK, differenziando tra chi ha subito bullismo e chi no. Nel primo caso, la distribuzione congiunta tra i due nodi è:

```
        AK
BB             0           1          99
  0 0.1583172 0.009527362 0.001979575
  1 0.5440760 0.031341708 0.007271546
  2 0.2315804 0.012549825 0.003356407
attr(,"class")
[1] "parray" "array"
```

Nel secondo:

```
        AK
BB              0            1           99
  0 0.39149449 0.019139728 0.004978332
  1 0.46232214 0.021783215 0.006266770
  2 0.08874528 0.003969356 0.001300688
attr(,"class")
[1] "parray" "array"
```

Tra chi subisce bullismo, la probabilità di avere un indice di distress alto e di aver subito delle multe per infrazioni al codice stradale è tre volte più elevata rispetto a coloro che non hanno subito bullismo (0,0125 vs. 0,0039). D'altra parte, la condizione "sana" di coloro che hanno un distress basso e non hanno mai subito multe, tra chi ha subito bullismo ha probabilità di avverarsi pari allo 0.16, tra chi non lo ha subito è più che doppia, pari allo 0.39. In base a quanto visto, se per un certo numero di studenti si conoscesse il punteggio del K6 e, in aggiunta, si avessero informazioni sull'eventuale bullismo subito, si potrebbero prevedere le probabilità con cui gli stessi potrebbero incorrere in multe per infrazioni al codice della strada.

Per avere, però, una previsione prudente, è necessario tener conto anche dell'incertezza. Coerentemente all'approccio bayesiano, per far questo si ricorre al metodo Monte Carlo che, simulando un numero elevato di casi, partendo dalle distribuzioni di probabilità congiunte, è in grado di fornire informazioni rispetto all'intervallo di valori più probabili entro cui cadrà la previsione. Per far questo si utilizza il pacchetto R MCMCpack (Andrew et al., 2011).

Partendo dall'ultimo esempio, si vuole valutare l'intervallo di credibilità al 95% entro cui cadranno le probabilità congiunte di BB e AK, condizionatamente al sottoinsieme di studenti che hanno subito bullismo. Come a-priori per la stima del valore dei parametri si considera una distribuzione Dirichlet non informativa[12]:

```
Iterations = 1:10000
Thinning interval = 1
Number of chains = 4
Sample size per chain = 10000

1. Empirical mean and standard deviation for each variable,
   plus standard error of the mean:

        Mean       SD  Naive SE Time-series SE
pi.1 0.151035 0.013689 6.845e-05      6.885e-05
pi.2 0.534058 0.019159 9.579e-05      9.687e-05
pi.3 0.224639 0.015958 7.979e-05      7.979e-05
pi.4 0.013834 0.004506 2.253e-05      2.248e-05
pi.5 0.037423 0.007304 3.652e-05      3.652e-05
pi.6 0.018210 0.005090 2.545e-05      2.565e-05
pi.7 0.006423 0.003082 1.541e-05      1.535e-05
```

---

[12] Il valore della a-priori della Dirichlet non informativa, come si può evincere dal secondo elemento dalla funzione MCmultinomdirichlet(), è pari a 10/27, in quanto il numeratore è rappresentato dal valore dell'iperparametro utilizzato nella determinazione della rete, il denominatore dal prodotto dei livelli dei nodi genitori di AK per i livelli del nodo BB.

```
pi.8 0.010889 0.003987 1.993e-05      2.004e-05
pi.9 0.003488 0.002274 1.137e-05      1.131e-05

2. Quantiles for each variable:

         2.5%      25%      50%      75%     97.5%
pi.1 0.1251394 0.141543 0.150629 0.160161 0.178572
pi.2 0.4962002 0.521229 0.534024 0.546977 0.571514
pi.3 0.1940320 0.213623 0.224459 0.235302 0.256304
pi.4 0.0064943 0.010570 0.013360 0.016609 0.023922
pi.5 0.0245155 0.032304 0.036915 0.042027 0.053006
pi.6 0.0096551 0.014557 0.017783 0.021375 0.029316
pi.7 0.0018447 0.004193 0.005942 0.008113 0.013834
pi.8 0.0045080 0.008008 0.010411 0.013274 0.019987
pi.9 0.0005503 0.001820 0.003007 0.004623 0.009124

> HPDinterval(simulazione,prob=0.95)
[[1]]
          lower        upper
pi.1 0.1242485967 0.177861427
pi.2 0.4984545367 0.572863684
pi.3 0.1943176195 0.256644349
pi.4 0.0053588596 0.022338370
pi.5 0.0240697920 0.052008516
pi.6 0.0091400967 0.028575231
pi.7 0.0012863520 0.012563265
pi.8 0.0037584125 0.018796462
pi.9 0.0002009738 0.007983574
attr(,"Probability")
[1] 0.95

[[2]]
          lower        upper
pi.1 0.1250980044 0.177247140
pi.2 0.4961607706 0.571824521
pi.3 0.1933636654 0.255812412
pi.4 0.0056506651 0.022593585
pi.5 0.0237466718 0.052091667
pi.6 0.0091290648 0.028758674
pi.7 0.0012319606 0.012476696
pi.8 0.0038848654 0.018687300
pi.9 0.0001883361 0.007855822
attr(,"Probability")
[1] 0.95

[[3]]
          lower        upper
pi.1 0.1237418473 0.177483962
pi.2 0.4965428236 0.571339096
pi.3 0.1928475240 0.254915279
pi.4 0.0056712486 0.022837741
pi.5 0.0236383689 0.051771106
pi.6 0.0090591024 0.028399540
pi.7 0.0013606415 0.012603438
pi.8 0.0039433632 0.019133016
pi.9 0.0001315425 0.007789573
attr(,"Probability")
[1] 0.95

[[4]]
          lower       upper
pi.1 0.1250128178 0.17837685
pi.2 0.4947625814 0.57003211
```

```
pi.3 0.1939274381 0.25563256
pi.4 0.0060276873 0.02311765
pi.5 0.0235056494 0.05192460
pi.6 0.0090722942 0.02822679
pi.7 0.0014234699 0.01267837
pi.8 0.0038689644 0.01862907
pi.9 0.0001432411 0.00804800
attr(,"Probability")
[1] 0.95
```

Grazie a questo passaggio, è possibile affermare che, con credibilità del 95%, la probabilità, per chi ha subito bullismo ed ha un indice di distress elevato, di subire multe è compresa tra 0,009 e 0,028 (rispetto allo 0,0125 puntuale visto in precedenza). La probabilità per chi ha subito bullismo ed ha un indice di distress basso, di non subire multe è compresa, con credibilità del 95%, tra 0,124 e 0,178 (l'indicazione puntuale precedente era pari a 0.158).

Con riferimento al sottoinsieme di chi non ha subito bullismo, si ottiene:

```
Iterations = 1:10000
Thinning interval = 1
Number of chains = 4
Sample size per chain = 10000

1. Empirical mean and standard deviation for each variable,
   plus standard error of the mean:

          Mean         SD   Naive SE  Time-series SE
pi.1 0.3946189 0.0090083 4.504e-05       4.529e-05
pi.2 0.4601800 0.0091628 4.581e-05       4.615e-05
pi.3 0.0910267 0.0052877 2.644e-05       2.644e-05
pi.4 0.0166797 0.0023532 1.177e-05       1.196e-05
pi.5 0.0241460 0.0028433 1.422e-05       1.413e-05
pi.6 0.0024931 0.0009138 4.569e-06       4.570e-06
pi.7 0.0045086 0.0012277 6.139e-06       6.197e-06
pi.8 0.0058863 0.0014081 7.040e-06       7.067e-06
pi.9 0.0004607 0.0003951 1.975e-06       1.958e-06

2. Quantiles for each variable:

          2.5%         25%         50%         75%       97.5%
pi.1 3.772e-01  0.3885307  0.3945840  0.4005983  0.412530
pi.2 4.423e-01  0.4539288  0.4602355  0.4663870  0.478133
pi.3 8.093e-02  0.0874434  0.0909187  0.0945246  0.101588
pi.4 1.238e-02  0.0150479  0.0165688  0.0181947  0.021587
pi.5 1.889e-02  0.0221709  0.0240424  0.0260126  0.030016
pi.6 1.035e-03  0.0018352  0.0023806  0.0030318  0.004582
pi.7 2.422e-03  0.0036331  0.0044062  0.0052627  0.007198
pi.8 3.455e-03  0.0048848  0.0057732  0.0067660  0.008927
pi.9 2.792e-05  0.0001769  0.0003544  0.0006309  0.001493

> HPDinterval(simulazione,prob=0.95)
[[1]]
            lower         upper
pi.1 3.775163e-01  0.412544790
pi.2 4.421429e-01  0.477806696
pi.3 8.092265e-02  0.101580374
pi.4 1.232979e-02  0.021441987
pi.5 1.857597e-02  0.029695358
pi.6 8.698568e-04  0.004278521
pi.7 2.325092e-03  0.007033016
```

```
pi.8 3.319736e-03 0.008676055
pi.9 3.545035e-07 0.001239208
attr(,"Probability")
[1] 0.95

[[2]]
           lower      upper
pi.1 3.771842e-01 0.412465398
pi.2 4.426539e-01 0.477805742
pi.3 8.065003e-02 0.101192514
pi.4 1.202748e-02 0.021185361
pi.5 1.863993e-02 0.029708244
pi.6 9.201072e-04 0.004347328
pi.7 2.322552e-03 0.007024864
pi.8 3.232595e-03 0.008631161
pi.9 1.301425e-06 0.001244543
attr(,"Probability")
[1] 0.95

[[3]]
           lower      upper
pi.1 3.776615e-01 0.412970121
pi.2 4.421219e-01 0.478340130
pi.3 8.076667e-02 0.101461818
pi.4 1.215564e-02 0.021341494
pi.5 1.880518e-02 0.029877952
pi.6 8.707618e-04 0.004265808
pi.7 2.248238e-03 0.006867911
pi.8 3.227217e-03 0.008656721
pi.9 9.572350e-07 0.001226283
attr(,"Probability")
[1] 0.95

[[4]]
           lower      upper
pi.1 3.765361e-01 0.411669796
pi.2 4.423714e-01 0.478022556
pi.3 8.079239e-02 0.101324115
pi.4 1.210158e-02 0.021209568
pi.5 1.882844e-02 0.029808589
pi.6 9.131772e-04 0.004334438
pi.7 2.300377e-03 0.007012418
pi.8 3.195412e-03 0.008605325
pi.9 2.284678e-07 0.001237039
attr(,"Probability")
[1] 0.95
```

Il giovane che non ha subito bullismo e con un indice di distress alto, ha una probabilità di subire multe che, con credibilità al 95%, sarà compresa tra 0,001 e 0,0043. Invece, un adolescente che non ha subito bullismo e con indice di distress basso, non subirà multe con probabilità compresa tra 0.38 e 0.41.

La rete, così come costruita, può essere utilizzata nello stimare le probabilità per un giovane, con determinate caratteristiche, di avere comportamenti di guida azzardati e pericolosi. Conoscendo la probabilità relativa alla transizione tra comportamenti rischiosi alla guida e futuri incidenti, anche elicitata esogenamente al modello sulla base del sapere di esperti, la rete potrebbe essere utilizzata per prevedere i possibili incidenti futuri tra giovani adolescenti.

**4.4 Possibili sviluppi futuri**

Dopo verifica da parte dell'esperto e con l'aggiunta di eventuali nodi/variabili che, allo stato attuale, risultano non osservati, la rete probabilistica presentata in questa sede potrebbe essere estesa e diventare una rete causale. La rete causale, a sua volta, potrebbe essere elaborata in rete decisionale, utile per guidare il governo regionale e le istituzioni preposte nella lotta al fenomeno in analisi.
Una rete siffatta implicherebbe che il policy maker possa intervenire almeno su alcuni nodi. Essendo, questi, legati da relazioni di causazione con i nodi figli, l'effetto sarebbe quello di riuscire a influenzare le probabilità delle modalità delle variabili a valle.
Ad ognuno di questi nodi andrebbe, inoltre, collegata una tabella di costi/benefici. In questa maniera, potendo il policy maker intervenire su un certo nodo, imponendone il valore, si avrebbe un'immediata quantificazione del costo di quel dato intervento e, grazie alla rete, si potrebbe stimare come i nodi discendenti saranno influenzati da quella modifica. Al termine dell'elaborazione complessiva, quando l'effetto si sarà propagato lungo i rami della rete interessati, sarà possibile valutare se quel dato intervento sia stato tale da produrre il beneficio massimo atteso, avendo tenuto conto dell'incertezza presente.

**CONCLUSIONI**

Il modello qui presentato prova ad affrontare un problema di primaria importanza che affligge la Toscana e non solo, qual è quello degli incidenti stradali degli adolescenti. In tal senso fondamentale è l'utilizzo della banca dati Edit 2015, per la ricchezza e la varietà delle informazioni raccolte. Sono proprio questi ultimi gli elementi che rendono utile il ricorso all'apprendimento automatico di reti bayesiane. Pensare di gestire decine di variabili, dal contenuto così diverso, in un modello da costruirsi manualmente, sarebbe un'impresa. Condurre la macchina in questa operazione, limitandosi a indicazioni di massima, o alla validazione di reti diverse prodotte dagli algoritmi di apprendimento, è operazione decisamente meno dispendiosa e più "umana". Evidenti sono, quindi, le potenzialità di questo tipo di strumenti in ambiti complessi quali quelli dei fenomeni sociali e psicologici.
Il modello presentato fornisce informazioni ulteriori rispetto a quelle dettagliate, ma più prettamente descrittive, presenti nel Rapporto dell'Osservatorio di Epidemiologia sui comportamenti rischiosi degli adolescenti alla guida. Una descrizione o una analisi di regressione logistica, come quelle presentate nel report, infatti, non sono in grado di cogliere le interrelazioni tra variabili, anche apparentemente "distanti" tra loro, come, invece, riesce a fare una modello a rete.
Interessante notare come, di massima, la rete qui realizzata sembri confermare quanto affermato da Elvik e Lund (2015). Come visto nel paragrafo 1.2, i due autori sottolineano come tassi di incidenti stradali più elevati si registrano tra i giovani maschi (nella rete, il nodo del genere è direttamente collegato alla tripletta di nodi riferiti a comportamenti rischiosi alla guida, in particolare al nodo relativo a guida sotto effetto di stupefacenti). In generale, la maggior probabilità dei giovani di avere incidenti stradali sarebbe imputabile a fattori biologici (ormoni, sviluppo celebrale), all'eccessiva autostima, ed in un atteggiamento di ribellione e di sfida tipici di questa età, atteggiamenti che possono essere per certi versi rappresentati dal nodo relativo al bullismo fatto, che si lega al nodo delle multe subite, e a quello della frequenza dell'uso di cannabis, che può essere considerato spia di un atteggiamento caratterizzato da sottovalutazione dei rischi (si può affermare che il consumatore di droghe, leggere o meno che siano, ha comunque scarso interesse per quelle che potrebbero essere eventuali effetti di tale comportamento sulla sua salute).
Poco è stato, qui, fatto per quel che concerne la potenziale conversione della rete da prettamente probabilistica a causale. Questo eventuale passaggio permetterebbe al modello di poter essere utile non solo in termini predittivi, trasformandolo in una sorta di vero e proprio "simulatore" di policy, con relativa valutazione di costi benefici, nelle mani del decisore.

Si può ritenere che, grazie alla Data Science, un domani, attraverso una prevenzione più mirata, si riuscirà a porre un minimo freno alla disinvoltura dei giovani alla guida.

**BIBLIOGRAFIA E SITOGRAFIA**

Andrew D. M., Kevin M. Q., & Jong H. P. (2011). MCMCpack: Markov Chain Monte Carlo in R. Journal of Statistical Software. 42(9): 1-21. URL http://www.jstatsoft.org/v42/i09/.

Bayes, T. & Price, R. (1763). "An Essay towards solving a Problem in the Doctrine of Chance. By the late Rev. Mr. Bayes, communicated by Mr. Price, in a letter to John Canton, A. M. F. R. S." (PDF). Philosophical Transactions of the Royal Society of London. 53 (0): 370–418. doi:10.1098/rstl.1763.0053.

Chickering, D. M. (2002). Learning equivalence classes of Bayesian-network structures. Journal of machine learning research, 2(Feb), 445-498.

Domingos, P. (2015). The master algorithm: How the quest for the ultimate learning machine will remake our world. Basic Books.

Elvik, R., & Lund, J. (2015). Contributing factors to traffic injuries in adolescents. The European Journal of Public Health, 25(suppl 3), ckv167-061.

Hansen K.D., Gentry J., Long L., Gentleman R., Falcon S., Hahne F., & Sarkar D. (2016). Rgraphviz: Provides plotting capabilities for R graph objects. R package version 2.18.0.

Heckerman D., Geiger D., & Chickering D. M. (1995). Learning Bayesian networks: The combination of knowledge and statistical data. Machine learning, 20(3), 197-243.

Højsgaard S. (2012). Graphical Independence Networks with the gRain Package for R. Journal of Statistical Software, 46(10), 1-26. URL http://www.jstatsoft.org/v46/i10/.

Istat (2015). Incidenti stradali – anno 2014 (http://www.istat.it/it/files/2015/11/Incidenti-stradali2014.pdf?title=Incidenti+stradali+in+Italia+-+03%2Fnov%2F2015+-+Testo+integrale.pdf).

Korb, K. B., & Nicholson, A. E. (2010). Bayesian artificial intelligence. CRC press.

Marchetti G. M., Drton M., & Sadeghi K. (2015). ggm: Functions for graphical Markov models. R package version 2.3. https://CRAN.R-project.org/package=ggm

Ministero delle Infrastrutture e dei Trasporti (2015), Statistiche sull'incidentalità dei trasporti stradali anche con riferimento alla tipologia di strada - Costi sociali dell'incidentalità stradale
(http://www.mit.gov.it/mit/site.php?p=cm&o=vd&id=4222).

Neapolitan, Richard E. (1989). Probabilistic reasoning in expert systems: theory and algorithms. Wiley. ISBN 978-0-471-61840-9.

Osservatorio di Epidemiologia, a cura di (2016). Comportamenti alla guida e stili di vita a rischio nei ragazzi in Toscana (https://www.ars.toscana.it/files/edit/2015/sintesi_EDIT_2015_def.pdf).

Pearl J. (2009). Causal inference in statistics: An overview. Statist. Surv. Volume 3 (2009), 96-146.

Pearl, J. (2003). Statistics and causal inference: A review. Test, 12(2), 281-345.

Pearl, J. (1988). Probabilistic Reasoning in Intelligent Systems. San Mateo, CA: Kaufmann, 23, 33-34.

Pearl, J. (1985). Bayesian Networks: A Model of Self-Activated Memory for Evidential Reasoning (UCLA Technical Report CSD-850017). Proceedings of the 7th Conference of the Cognitive Science Society, University of California, Irvine, CA. pp. 329–334. Retrieved 2009-05-01.

Pearl, J., Glymour, M., & Jewell, N. P. (2016). Causal Inference in Statistics: A Primer. John Wiley & Sons.

Rahim Khan, U., Sengoelge, M., Zia, N., A Razzak, J., Hasselberg, M., & Laflamme, L. (2015). Global time differences in road traffic injuries among children and adolescents between and 1990 and 2013: Regional and economical perspectives from global burden of diseases study. Journal of Local and Global Heath Science, (2), 62.

Scutari M., & Denis J. P. (2014). Bayesian Networks with Examples in R. Chapman and Hall, Boca Raton. ISBN 978-1482225587.
World Health Organization (2014). Health for The World's Adolescence: A Second Chance in The Second Decade (http://apps.who.int/adolescent/second-decade/).